\documentclass[reprint, amsmath,amssymb, aps, prl, longbibliography, lengthcheck]{revtex4}

\usepackage{graphicx}
\usepackage{dcolumn}
\usepackage{bm}
\usepackage{hyperref}

\usepackage{pstricks,pst-3d,pst-text,pst-node}

\newcommand{\student}[1]{{#1}}

\def\nten#1{\mbox{\boldmath{$\rm #1$}}}
\def\nvec#1{\mbox{\boldmath{$\rm #1$}}}

\def\Dp{\nten D^{\rm p}}
\def\nv{\nvec n^{(3)}}

\newpsobject{showgrid}{psgrid}{subgriddiv=1,griddots=10,gridlabels=10pt}

\begin{document}

\preprint{APS/123-QED}

\title{A Mechanism of Failure in Shear Bands}

\author{Mohammad E. Torki$^{1,2}$}

\author{A. Amine Benzerga$^{1,2,3}$}
\email{benzerga@tamu.edu}

\affiliation{$^1$ Department of Aerospace Engineering, Texas A\&M University, 
             College Station, TX 77843, USA}
\affiliation{$^2$ Center for intelligent Multifunctional Materials and Structures, 
	     TEES, College Station, TX 77843, USA}
\affiliation{$^3$ Department of Materials Science and Engineering, Texas A\&M University, 
             College Station, TX 77843, USA}

\date{\today}

\begin{abstract}
We have carried out dilatant plasticity simulations to investigate the process
of void-mediated failure inside a shear band. The constitutive model accounts for
possibly inhomogeneous flow within the band, void rotation and void elongation.
We found that the material in the band may soften with no increase
in the void volume fraction. 
For a given matrix hardening capacity, the rate of softening was found to depend
strongly on the ratio of shear band width to in-plane void spacing.
The emergent softening led to complete loss of load bearing capacity thereby
providing a physical mechanism of failure in shear bands.
The mechanism is consistent with essential features of shear-fractured
specimens in terms of surface roughness, porosity and dimple shape.
\end{abstract}

\maketitle

Failure by shear banding is ubiquitous and occurs
in complex fluids \cite{Schall10},
granular materials \cite{Alshibli00,Gourlay07},
rocks \cite{Gomez12}
polycrystals \cite{Ma02,Morgeneyer11b},
polymers \cite{Friedrich83} and
metallic glasses \cite{Greer13,Hofmann08}.
However, the mechanism of material separation inside the shear band has remained elusive.
Understanding it will not only potentially retard failure in shear bands, if desired,
but also impact other applications where failure occurs under shear dominated loadings.
The stress state in shear bands is generally complex depending on the loading path
prior to the onset of strain localization \cite{Morgeneyer16}.
Correspondingly, shear bands are generally dilational.
While arbitrarily large tension-to-shear ratios may be encountered inside shear bands,
here we focus on situations of vanishingly small tension-to-shear ratios
and aim to present a physical model of complete material separation.

\begin{figure}[t]
	\[ \includegraphics[width=0.85\columnwidth]{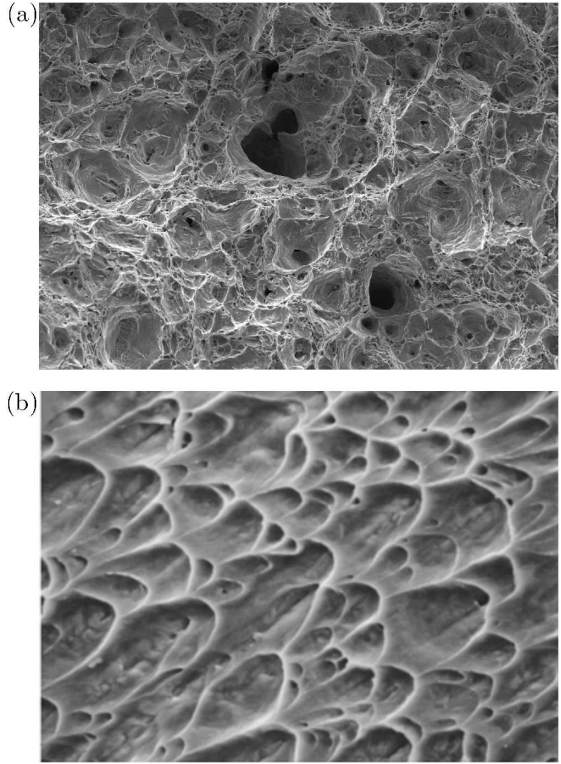} \]
	\caption{Typical fracture surfaces of metals failing in 
	(a) tension, and (b) shear \cite{PineauAM}.}
	\label{fig:fracto}
\end{figure}

Voids are the main defects mediating ductile fracture \cite{BenzergaAAM,PineauAM}.
The plastic enlargement of these defects dominates at moderate to high
ratios of tension-to-shear stress (tension-dominated loading), Fig.~\ref{fig:fracto}a.
Voids are also believed to play an important role at low tension-to-shear ratios
(shear-dominated loading), Fig.~\ref{fig:fracto}b.
However, a specific mechanism by which failure occurs is still lacking.
Void nucleation is material specific and will not be addressed here.
Well-established micromechanical models of void growth and coalescence 
\cite{Gurson77,Tvergaard84} predict infinite ductility under shear loading.
This is due to two idealizations: (i) that the void volume fraction, $f$, is the sole internal
parameter representing the defects; and (ii) that void coalescence occurs upon
attainment of a critical value of $f$. 
Since the rate of growth of $f$ is completely determined by the dilational part of
the macroscopic plastic strain rate, which is nil in shear, no growth is predicted,
hence no failure.
An attempt to remedy this consists of amending the void growth law with a shear-dependent
term \cite{Nahshon08}. This proposal is attractive but presents two shortcomings.
Not only does it violate the principle of mass conservation underlying the void growth law
but it also presumes that a void-growth-like behavior is required for failure in shear.
On the other hand, a much earlier mechanism-based model limited to isolated voids \cite{McClintock66}
did highlight the essential role of void
rotation and possible linkage of neighboring voids by mere impingement.
More recent direct numerical simulations
\cite{Tvergaard08,Tvergaard09,Nielsen12} 
revealed the existence of a maximum
in the shear load response and exhibited three essential microscopic features:
(i) void-induced strain localization at the sub-cell level;
(ii) void rotation; and
(iii) void elongation in the rotated state.
However, such calculations are extremely challenging, and thus cannot be pursued
much beyond the maximum load. Furthermore, they are not scalable
so that a coarse-grained continuum model that
mimics the behavior gleaned from these simulations is lacking
\cite{Benzerga16}.

Quite recently, Morin et al. \cite{Morin16} proposed mechanism-based
modeling of failure under shear-dominated loadings. Their model accounts
for void rotation and void shape change,
but fails to account
for the void-induced strain localization that occurs from the outset in shear.
It also employs an {\it ad hoc} coalescence criterion, reminiscent of the critical
void volume criterion used in conjunction with the Gurson model \cite{Tvergaard84}.
In this Letter, we present a parameter-free model of failure under shear-dominated
loading, which accounts for sub-cell strain localization, void rotation
and void shape change, and discuss the model's capabilities to simulate
complete loss of stress carrying capacity in shear.

\begin{figure}[h]
	\includegraphics[width=\columnwidth]{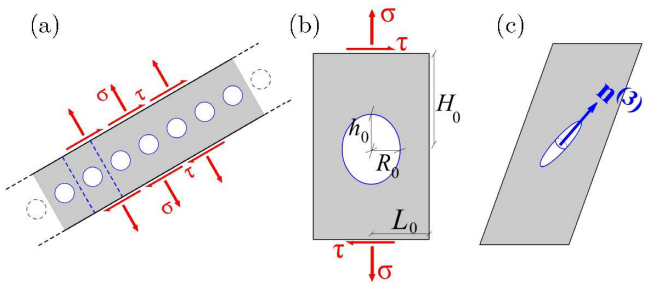}
	\caption{Problem formulation.
	(a) Doubly periodic row of voids inside the shear band.
	(b) Geometry of undeformed elementary cell.
	(c) Homogeneous deformation of the cell involving void rotation.}
	\label{fig:pb-sketch}
\end{figure}

\begin{figure}[h]
	\includegraphics[width=\columnwidth]{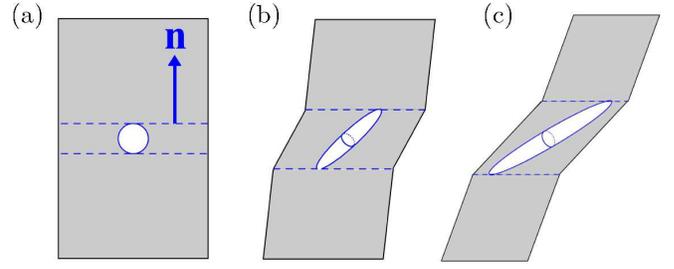}
	\caption{Essential features captured by the coarse-grained model.
	(a) Void-induced strain localization.
	(b) Void rotation.
	(c) Void elongation.}
	\label{fig:pb-sketch2}
\end{figure}

When voids are at the micron scale and above, void-mediated fracture 
in the shear band may be described by continuum mechanics.
The shear band is assumed to be acted upon by
a shear stress, $\tau$, and a normal tensile stress, $\sigma$, Fig.~\ref{fig:pb-sketch}a.
Voids are assumed to have nucleated, in some way, inside the band.
A regular doubly-periodic array of voids is assumed for simplicity
so that analysis of a single tetragonal cell, Fig.~\ref{fig:pb-sketch}b, is sufficient.
The aspect ratio of the cell, $\lambda \equiv H/L$, represents the current ratio of shear band thickness
to in-plane void spacing. Other cell parameters include the void volume fraction, $f$,
the void aspect ratio, $w \equiv h/R$, 
along with two unit vectors: $\nvec n^{(3)}$ for the orientation of the void, modeled as a spheroid
when it deforms, Fig.~\ref{fig:pb-sketch}c,
and $\nvec n$ for the orientation of the localization plane, Fig.~\ref{fig:pb-sketch2}.
Multiple possibilities for $\nvec n$ may be chosen
depending on the underlying spatial arrangement of voids. Here, only one such orientation
is considered, which is normal to the shear band. Initial values of all internal
parameters are indicated with subscript 0.

To investigate failure in shear, we carried out 
numerical simulations using a continuum micromechanics framework
for dilatant plasticity that captures the essential features of sub-cell
deformation sketched in Fig.~\ref{fig:pb-sketch2}.
Contrary to the direct numerical simulations in \cite{Tvergaard08,Tvergaard09,Nielsen12},
we do not model the void explicitly, but through the coarse-grained model.
The effective elasticity domain is represented by the intersection of two convex domains.
Its boundary is therefore determined by the intersection of two surfaces in stress space
$\Phi^{\rm H} (\nten\sigma; f,w,\nvec n^{(3)}) = 0$ and 
$\Phi^{\rm I} (\nten\sigma; f,w, \lambda, \nvec n^{(3)}, \nvec n) = 0$
where yield functions $\Phi^{\rm H}$ and $\Phi^{\rm I}$ correspond to 
homogeneous (Fig.~\ref{fig:pb-sketch}c) and 
inhomogeneous (Fig.~\ref{fig:pb-sketch2}b) deformation of the cell, respectively.
The plastic portion of the symmetric part of the velocity gradient, 
$\nabla \nvec v \equiv {\nten L}$, is obtained
by normality to the effective yield surface.
For generally tensile stress states $\nten\sigma$, there is competition between the two yielding
mechanisms \cite{BenzergaJMPS,Benzerga04AM-II} with $\Phi^{\rm H} =0$ prevailing in
the early stages of any triaxial stressing process.
For combined tension and shear, as in Fig.~\ref{fig:pb-sketch}b, it is the $\sigma/\tau$ ratio 
that determines which yielding mechanism would dominate.
Yield functions derived from first principles of micromechanics are used for $\Phi^{\rm H}$
\cite{KeralavarmaJMPS} and $\Phi^{\rm I}$ \cite{BenzergaJAM,TorkiJAM}.
Evolution equations for the internal parameters in $\Phi^{\rm H}$ were derived in 
\cite{KeralavarmaJMPS}; also see \cite{KweonCMAME} for computational details.
For sufficiently low $\sigma/\tau$ (shear-dominated loading) inhomogeneous yielding
dominates from the outset so that all subsequent deformation history is governed by $\Phi^{\rm I}$.
Details about the two-surface formulation may be found as Supplemental Material.
Here, it suffices to exhibit the governing equations for inhomogeneous yielding:
\def\Svol{{\cal V}}
\def\Ssurf{{\cal S}}
\def\Shear{{(1-\chib^2)\bar\tau}}
\def\chib{\bar\chi}
\def\wb{\bar w}
\def\tbar{\bar\tau}
\def\sbar{\bar\sigma}
\begin{equation}
\Phi^{\rm I} =
\begin{cases}\displaystyle
	\left(\frac{|\sigma| - \Ssurf (\chib,\wb)}{\Svol(\chib)} \right)^2 + \frac{\tau^2}{\Shear^2} - 1
 & \mbox{for} \, |\sigma| \ge \Ssurf
\\ \\ \displaystyle
 \frac{\tau^2}{\Shear^2} - 1
 & \mbox{for} \, |\sigma| \le \Ssurf
\end{cases}
\label{eq:TBL}
\end{equation}
with
\begin{equation}
\Svol/\tbar =  2 - \sqrt{1+3\chib^4} + \ln \frac{1 + \sqrt{1+3\chib^4}}{3 \chib^2} 
\label{eq:Svol}
\end{equation}
\begin{equation}
\Ssurf / \tbar = \frac{\chib^3 - 3 \chib + 2}{3 \chib \wb} 
\label{eq:Ssurf}
\end{equation}
where the effective ligament parameter, $\chib$, and effective void aspect ratio, $\wb$,
correspond to an equivalent cylindrical void with axis $\nvec n$, 
obtained by a volume-preserving projection of the rotating spheroidal void 
onto the localization plane. The exact shape, spheroidal versus cylindrical, has little incidence
on yielding \cite{MorinJMPS}. However,
since equations~(\ref{eq:TBL}) were derived for cylindrical voids \cite{TorkiJAM},
this choice is made here (see Supplemental Material).
Also, $\tbar$ is the flow stress in shear of the material without voids,
taken as a power law in the effective plastic strain
$\tbar = \tau_0 (1+ E \bar\gamma / 3\tau_0)^N$ with
$\tau_0$ the initial yield strength, $E$ Young's modulus, and $N$ the hardening exponent.
Implicit dependence upon the void axis $\nvec n^{(3)}$ in~(\ref{eq:TBL}) is through $w$
and dependence upon the localization plane normal is through $\chib$, $\wb$ as well as
$\sigma = \nvec n \cdot \nten\sigma \nvec n$ and $\tau = \nvec m \cdot \nten\sigma \nvec n$
with $\nvec m$ a unit vector along the applied shear.

Upon continued plastic loading, the structure evolves according to
(with $C= \nv \cdot \nvec n$, $S= \nv \cdot \nvec m$ and
$c^3 =  C^3 {(fw^2)}/{\lambda^2}$):
\begin{equation}
\dot f = (1-f) D^{\rm p}_{kk} = (1-f) \Lambda \frac{\partial \Phi^{\rm I}}{\partial \sigma}
\label{eq:fdot}
\end{equation}
\begin{equation}
	\frac{\dot w}{w} = \frac{C}{2 c} \left(\left[3C - \frac{1}{\chib^2} \left(\frac{w}{\wb} \right)^\frac23 \right] \nvec n \cdot \Dp \nvec n
	+ 6 S \nvec m \cdot \Dp \nvec n \right) 
\label{eq:wdot}
\end{equation}
\begin{equation}
	\dot{\nvec n}^{(3)} =  (\nten \Omega^v + \nten \Omega^l )  \nv
\label{eq:n3dot}
\end{equation}
\begin{equation}
	{\small
	\nten \Omega^l =  \left( \frac{\dot c}{c} - \frac13 \left[\frac{\dot f}{f} + 2 (\frac{\dot w}{w} - \frac{\dot \lambda}{\lambda}) \right] \right) 
	\left( \nvec n \otimes \nvec n - \frac{C^2}{S^2} \nvec m \otimes \nvec m  \right)
	}
\label{eq:Omega_l}
\end{equation}
\begin{equation}
	\dot c = (1-c) D_{33}
\label{eq:cdot}
\end{equation}
\begin{equation}
\dfrac{\lambda}{\lambda_0} = \dfrac{1}{\sqrt{\mathcal J}} \left( \nvec n \cdot \nten F \nten F^T \nvec n \right)^{\frac34}
\label{eq:lupdate}
\end{equation}
\begin{equation}
\nvec n = \dfrac{\nten F^{- \rm T} \nvec n_{(0)}}{|\nten F^{- \rm T} \nvec n_{(0)}|}
\label{eq:n}
\end{equation}
Eqn.~(\ref{eq:fdot}) expresses plastic incompressibility of the matrix \cite{Gurson77}
and Eqns.~(\ref{eq:wdot}) and~(\ref{eq:n3dot}) are for the constrained
motion of the top and bottom void boundaries due to elastic unloading
above and below the void
(see Supplemental Material for derivations.)
Also, $\Lambda$ in~(\ref{eq:fdot}) is the plastic multiplier,
$\Dp \equiv \mbox{sym} \nten L^{\rm p}$,
$\nten \Omega^v$ in~(\ref{eq:n3dot}) represents the deviation from
the continuum spin due to the eigen-rotation of the void, calculated
using Eshelby concentration tensors \cite{Eshelby57} after
\cite{Kailasam98,KweonCMAME}, $\nten \Omega^l$
is the shear-induced rotation that comes from mere distortion of void boundaries
(dominant here), and $\nten F$ is the deformation gradient used to update
the band orientation $\nvec n$.
Relations~(\ref{eq:wdot})--(\ref{eq:cdot}) are straightfroward generalizations of the evolution equations
of Benzerga \cite{BenzergaJMPS} in the absence of shear, whereas
(\ref{eq:lupdate}) is taken from \cite{Leblond08}.
The effective plastic strain is evolved using Gurson's identity:
\begin{equation}
\nten \sigma : {\nten L}^{\rm p} = (1-f) \tbar \dot{\bar\gamma}
\label{eq:powbalance}
\end{equation}
The above plastic relations were augmented with hypoelasticity
within an objective co-rotational finite deformation framework.
The nonlinear constitutive relations were integrated using
an implicit time integration scheme similar to \cite{KweonCMAME}.

\begin{figure}[h]
	\[ \includegraphics{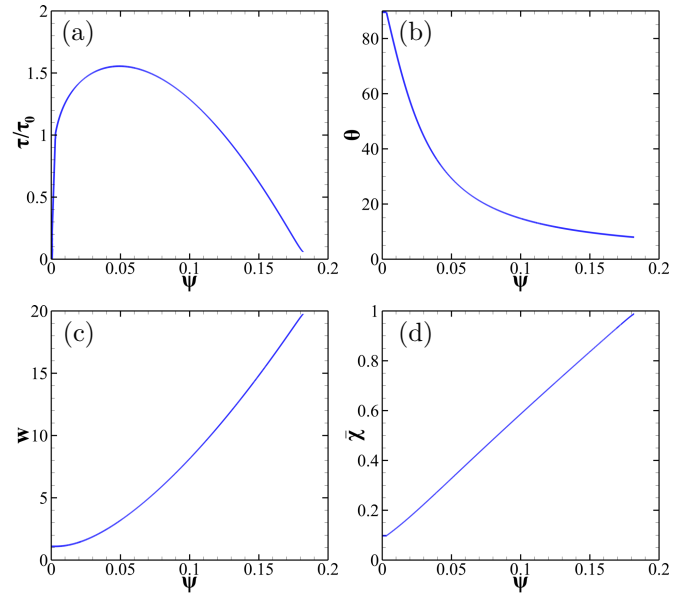} \]
	\caption{Typical results for vanishingly small tension-to-shear ratio, $\sigma/\tau$.
	Against the shear angle, $\psi$, are plotted the
	(a) shear stress, $\tau$, in units of $\tau_0$; (b) void orientation, $\theta$ ($^{\circ}$);
	(c) void aspect ratio, $w$; and (d) effective ligament parameter, $\chib$.
	Simulation parameters are: $f_0 = 0.0005$, $w_0=1.1$, $\lambda_0=2$, $N=0.2$, 
	$\sqrt 3 \tau_0 / E= 0.002$.}
	\label{fig:l2}
\end{figure}

The typical shear stress versus shear angle response, Fig.~\ref{fig:l2}a,
results from competing effects of matrix hardening (set by $N$)
and microstructural softening
induced by void rotation, Fig.~\ref{fig:l2}b, and elongation in the rotated state,
Fig.~\ref{fig:l2}c.
The angle $\theta$ is such that $\cos \theta = \nv \cdot \nvec n $.
As a result, the area of the void projected onto the plane of localization increases monotonically,
as captured through the effective ligament parameter, $\chib$, Fig.~\ref{fig:l2}d.
When $\chib$ approaches unity all stress carrying capacity vanishes by virtue of~(\ref{eq:TBL}).
This occurs while the void volume fraction $f$ remains constant (not shown).
In actuality, some decrease in $f$ is expected. To capture this detail would require 
a three-dimensional void model \cite{Morin16} and would have little effect on essential behavior
(see Supplemental Material for further details).

%

\begin{figure}[h]
	\[ \includegraphics[width=0.75\columnwidth]{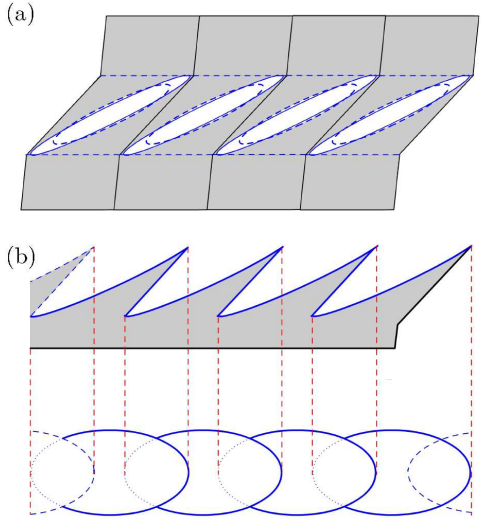} \]
	\caption{Predicted failure mechanism in shear and its connection to the fracture surface
	of Fig.~\ref{fig:fracto}b. 
	(a) Few neighboring cells near the ultimate state $\chib = 1$.
	(b) Side and top views of cut-out from (a) after material separation.}
	\label{fig:disc-sketch}
\end{figure}

To link the above findings with salient features of sheared fracture surfaces, Fig.~\ref{fig:fracto}b,
consider few neighboring cells at about the ultimate state $\chib = 1$ 
(dashed in Fig.~\ref{fig:disc-sketch}a).
We assume that final linkup would occur by some finer-scale microshear process, Fig.~\ref{fig:disc-sketch}b.
In reality, void distributions are not periodic so that it is likely that when the elongated
void reaches the lateral boundaries, it will link up with a neighboring void.
Details aside, a top view of the so-simulated fracture surface,
Fig.~\ref{fig:disc-sketch}c, provides a rationale for three key experimental observations:
(i) parabolic dimples;
(ii) low surface roughness; and
(iii) low local porosity, relative to tensile fracture surfaces, Fig.~\ref{fig:fracto}a.
Both roughness and porosity are related to the dimple height, which is set by the amount of rotation
prior to failure. In the example shown (Fig.~\ref{fig:l2}b) the rotation is actually much more
than depicted in Fig.~\ref{fig:disc-sketch}.

\begin{figure}[h]
	\[ \includegraphics{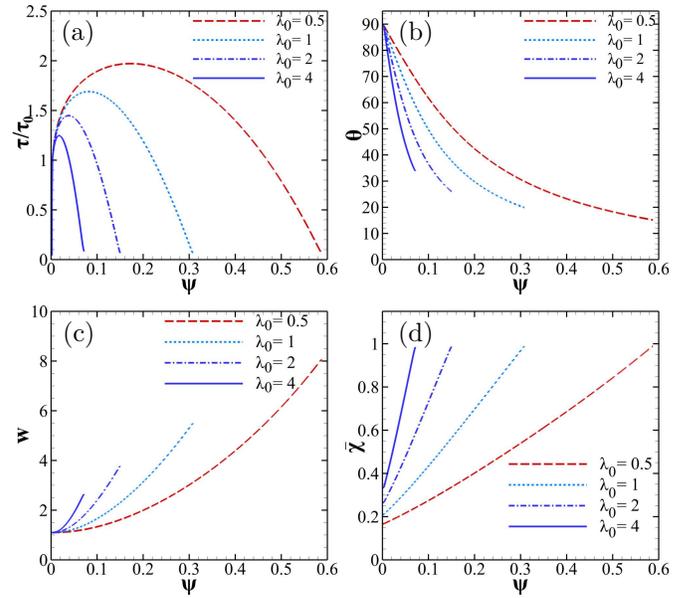} \]
	\caption{Effect of the ratio, $\lambda_0$, of shear-band width to in-plane void spacing
	for $f_0 = 0.01$, $w_0=1.1$, $N=0.2$, 
	$\sqrt 3 \tau_0 / E= 0.002$.}
	\label{fig:l}
\end{figure}

At fixed hardening capacity of the matrix material 
and fixed void volume fraction $f_0$, the strain to failure
is dependent upon the ratio $\lambda_0$ of initial band thickness to in-plane void spacing, Fig.~\ref{fig:l}a.
The rotation of the void, Fig.~\ref{fig:l}b, and its aspect ratio (Fig.~\ref{fig:l}c) are 
also sensitive to $\lambda_0$.
For fixed porosity of initially spherical voids ($w_0 \approx 1$),
varying $\lambda_0$ amounts to varying $\chib_0 = R_0/L_0$ (Fig.~\ref{fig:l}d), which has a direct effect on
the strain to failure.

\begin{figure}[h]
	\includegraphics{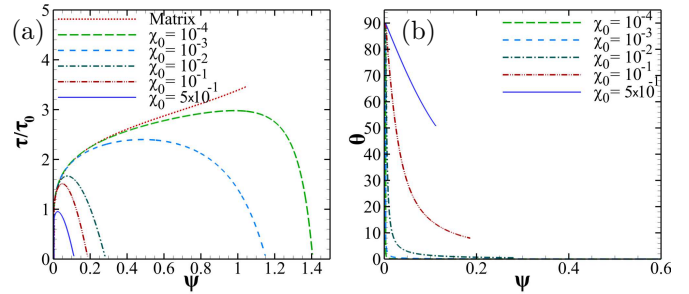}
	\caption{(a) Shear response and (b) evolution of void rotation for
	various values of the initial ligament parameter $\chib_0 \equiv R_0/L_0$
	and $w_0=1.1$, $\lambda_0=2$, $N=0.2$, and
	$\sqrt 3 \tau_0 / E= 0.002$.}
	\label{fig:chi0}
\end{figure}

To investigate this effect further, the initial ligament parameter $\chib_0$ was varied
over four decades. The results in Fig.~\ref{fig:chi0} show that the strain to failure
is essentially set by the value of $\chib_0$, which represents the size of voids that nucleate
inside the shear band, relative to their in-plane spacing.
In case of the dense matrix (no voids), the response becomes homogeneous as expected,
Fig.~\ref{fig:chi0}a. However, even in the dilute case $f_0 \to 0$ failure is predicted
so long as there are voids. Of course, there is a lower limit in a continuum description
on the void size implied by a given choice of $\chib_0 \to 0$. 
What is essential is that the lower the value of $\chib_0$ the more localized the
plastic flow inside the shear band, and this leads to a faster rotation of the voids,
as illustrated in Fig.~\ref{fig:chi0}b.
This result suggests that materials in which small voids are able to nucleate inside
the shear band, such as metallic glasses \cite{Shao14},
would have a lower fracture surface roughness. 
This is due to the vanishingly small dimple height, as the voids would
have almost completely rotated.

Our results illustrate for the first time a possible mechanism of failure inside shear bands.
They also show that a micromechanical dilatant plasticity framework can provide new insights
into aspects of material behavior heretofore not explained using either continuum or
atomistic approaches. The predictions connect macroscopic behavior
with detailed microscopic information about observables (the voids) and measurable
attributes thereof. Once the model is implemented to solve boundary-value
problems, further contact with experiments can be made.

We gratefully acknowledge financial support from the National Science Foundation 
under grant CMMI-1405226.


\begin{thebibliography}{32}
\expandafter\ifx\csname natexlab\endcsname\relax\def\natexlab#1{#1}\fi
\expandafter\ifx\csname bibnamefont\endcsname\relax
  \def\bibnamefont#1{#1}\fi
\expandafter\ifx\csname bibfnamefont\endcsname\relax
  \def\bibfnamefont#1{#1}\fi
\expandafter\ifx\csname citenamefont\endcsname\relax
  \def\citenamefont#1{#1}\fi
\expandafter\ifx\csname url\endcsname\relax
  \def\url#1{\texttt{#1}}\fi
\expandafter\ifx\csname urlprefix\endcsname\relax\def\urlprefix{URL }\fi
\providecommand{\bibinfo}[2]{#2}
\providecommand{\eprint}[2][]{\url{#2}}

\bibitem[{\citenamefont{Schall and van Hecke}({2010})}]{Schall10}
\bibinfo{author}{\bibfnamefont{P.}~\bibnamefont{Schall}} \bibnamefont{and}
  \bibinfo{author}{\bibfnamefont{M.}~\bibnamefont{van Hecke}},
  \bibinfo{journal}{{Annual Review of Fluid Mechanics}}
  \textbf{\bibinfo{volume}{{42}}}, \bibinfo{pages}{67}
  (\bibinfo{year}{{2010}}).

\bibitem[{\citenamefont{Alshibli and Sture}(2000)}]{Alshibli00}
\bibinfo{author}{\bibfnamefont{K.~A.} \bibnamefont{Alshibli}} \bibnamefont{and}
  \bibinfo{author}{\bibfnamefont{S.}~\bibnamefont{Sture}}, \bibinfo{journal}{J.
  Geotech. Geoenviron. Eng.} \textbf{\bibinfo{volume}{126}},
  \bibinfo{pages}{495} (\bibinfo{year}{2000}).

\bibitem[{\citenamefont{Gourlay and Dahle}({2007})}]{Gourlay07}
\bibinfo{author}{\bibfnamefont{C.~M.} \bibnamefont{Gourlay}} \bibnamefont{and}
  \bibinfo{author}{\bibfnamefont{A.~K.} \bibnamefont{Dahle}},
  \bibinfo{journal}{{Nature}} \textbf{\bibinfo{volume}{{445}}},
  \bibinfo{pages}{70} (\bibinfo{year}{{2007}}).

\bibitem[{\citenamefont{Gomez-Rivas and Griera}({2012})}]{Gomez12}
\bibinfo{author}{\bibfnamefont{E.}~\bibnamefont{Gomez-Rivas}} \bibnamefont{and}
  \bibinfo{author}{\bibfnamefont{A.}~\bibnamefont{Griera}},
  \bibinfo{journal}{{J. Struct. Geol.}} \textbf{\bibinfo{volume}{{34}}},
  \bibinfo{pages}{61} (\bibinfo{year}{{2012}}).

\bibitem[{\citenamefont{Wei et~al.}({2002})\citenamefont{Wei, Jia, Ramesh, and
  Ma}}]{Ma02}
\bibinfo{author}{\bibfnamefont{Q.}~\bibnamefont{Wei}},
  \bibinfo{author}{\bibfnamefont{D.}~\bibnamefont{Jia}},
  \bibinfo{author}{\bibfnamefont{K.~T.} \bibnamefont{Ramesh}},
  \bibnamefont{and} \bibinfo{author}{\bibfnamefont{E.}~\bibnamefont{Ma}},
  \bibinfo{journal}{Appl. Phys. Lett.} \textbf{\bibinfo{volume}{{81}}},
  \bibinfo{pages}{1240} (\bibinfo{year}{{2002}}).

\bibitem[{\citenamefont{Morgeneyer and Besson}(2011)}]{Morgeneyer11b}
\bibinfo{author}{\bibfnamefont{T.~F.} \bibnamefont{Morgeneyer}}
  \bibnamefont{and} \bibinfo{author}{\bibfnamefont{J.}~\bibnamefont{Besson}},
  \bibinfo{journal}{Scr. Mater.} \textbf{\bibinfo{volume}{65}},
  \bibinfo{pages}{1002} (\bibinfo{year}{2011}).

\bibitem[{\citenamefont{Friedrich}({1983})}]{Friedrich83}
\bibinfo{author}{\bibfnamefont{K.}~\bibnamefont{Friedrich}},
  \bibinfo{journal}{{Adv. Pol. Sci.}} \textbf{\bibinfo{volume}{{52-3}}},
  \bibinfo{pages}{225} (\bibinfo{year}{{1983}}).

\bibitem[{\citenamefont{Greer et~al.}({2013})\citenamefont{Greer, Cheng, and
  Ma}}]{Greer13}
\bibinfo{author}{\bibfnamefont{A.~L.} \bibnamefont{Greer}},
  \bibinfo{author}{\bibfnamefont{Y.~Q.} \bibnamefont{Cheng}}, \bibnamefont{and}
  \bibinfo{author}{\bibfnamefont{E.}~\bibnamefont{Ma}},
  \bibinfo{journal}{Mater. Sci. Eng.:R} \textbf{\bibinfo{volume}{{74}}},
  \bibinfo{pages}{71} (\bibinfo{year}{{2013}}).

\bibitem[{\citenamefont{Hofmann et~al.}({2008})\citenamefont{Hofmann, Suh,
  Wiest, Duan, Lind, Demetriou, and Johnson}}]{Hofmann08}
\bibinfo{author}{\bibfnamefont{D.~C.} \bibnamefont{Hofmann}},
  \bibinfo{author}{\bibfnamefont{J.-Y.} \bibnamefont{Suh}},
  \bibinfo{author}{\bibfnamefont{A.}~\bibnamefont{Wiest}},
  \bibinfo{author}{\bibfnamefont{G.}~\bibnamefont{Duan}},
  \bibinfo{author}{\bibfnamefont{M.-L.} \bibnamefont{Lind}},
  \bibinfo{author}{\bibfnamefont{M.~D.} \bibnamefont{Demetriou}},
  \bibnamefont{and} \bibinfo{author}{\bibfnamefont{W.~L.}
  \bibnamefont{Johnson}}, \bibinfo{journal}{{Nature}}
  \textbf{\bibinfo{volume}{{451}}}, \bibinfo{pages}{1085}
  (\bibinfo{year}{{2008}}).

\bibitem[{\citenamefont{Morgeneyer et~al.}(2016)\citenamefont{Morgeneyer,
  Taillandier-Thomas, Buljac, Helfen, and Hild}}]{Morgeneyer16}
\bibinfo{author}{\bibfnamefont{T.~F.} \bibnamefont{Morgeneyer}},
  \bibinfo{author}{\bibfnamefont{T.}~\bibnamefont{Taillandier-Thomas}},
  \bibinfo{author}{\bibfnamefont{A.}~\bibnamefont{Buljac}},
  \bibinfo{author}{\bibfnamefont{L.}~\bibnamefont{Helfen}}, \bibnamefont{and}
  \bibinfo{author}{\bibfnamefont{F.}~\bibnamefont{Hild}}, \bibinfo{journal}{J.
  Mech. Phys. Solids} \textbf{\bibinfo{volume}{96}}, \bibinfo{pages}{550}
  (\bibinfo{year}{2016}).

\bibitem[{\citenamefont{Benzerga and Leblond}(2010)}]{BenzergaAAM}
\bibinfo{author}{\bibfnamefont{A.~A.} \bibnamefont{Benzerga}} \bibnamefont{and}
  \bibinfo{author}{\bibfnamefont{J.-B.} \bibnamefont{Leblond}},
  \bibinfo{journal}{Adv. Appl. Mech.} \textbf{\bibinfo{volume}{44}},
  \bibinfo{pages}{169} (\bibinfo{year}{2010}).

\bibitem[{\citenamefont{Pineau et~al.}(2016)\citenamefont{Pineau, Benzerga, and
  Pardoen}}]{PineauAM}
\bibinfo{author}{\bibfnamefont{A.}~\bibnamefont{Pineau}},
  \bibinfo{author}{\bibfnamefont{A.~A.} \bibnamefont{Benzerga}},
  \bibnamefont{and} \bibinfo{author}{\bibfnamefont{T.}~\bibnamefont{Pardoen}},
  \bibinfo{journal}{Acta Mater.} \textbf{\bibinfo{volume}{107}},
  \bibinfo{pages}{424} (\bibinfo{year}{2016}).

\bibitem[{\citenamefont{Gurson}(1977)}]{Gurson77}
\bibinfo{author}{\bibfnamefont{A.~L.} \bibnamefont{Gurson}},
  \bibinfo{journal}{J. Eng. Mat. Tech.} \textbf{\bibinfo{volume}{99}},
  \bibinfo{pages}{2} (\bibinfo{year}{1977}).

\bibitem[{\citenamefont{Tvergaard and Needleman}(1984)}]{Tvergaard84}
\bibinfo{author}{\bibfnamefont{V.}~\bibnamefont{Tvergaard}} \bibnamefont{and}
  \bibinfo{author}{\bibfnamefont{A.}~\bibnamefont{Needleman}},
  \bibinfo{journal}{Acta Metall.} \textbf{\bibinfo{volume}{32}},
  \bibinfo{pages}{157} (\bibinfo{year}{1984}).

\bibitem[{\citenamefont{Nahshon and Hutchinson}(2008)}]{Nahshon08}
\bibinfo{author}{\bibfnamefont{K.}~\bibnamefont{Nahshon}} \bibnamefont{and}
  \bibinfo{author}{\bibfnamefont{J.~W.} \bibnamefont{Hutchinson}},
  \bibinfo{journal}{Eur. J. Mech. A} \textbf{\bibinfo{volume}{27}},
  \bibinfo{pages}{1} (\bibinfo{year}{2008}).

\bibitem[{\citenamefont{McClintock et~al.}(1966)\citenamefont{McClintock,
  Kaplan, and Berg}}]{McClintock66}
\bibinfo{author}{\bibfnamefont{F.~A.} \bibnamefont{McClintock}},
  \bibinfo{author}{\bibfnamefont{S.~M.} \bibnamefont{Kaplan}},
  \bibnamefont{and} \bibinfo{author}{\bibfnamefont{C.~A.} \bibnamefont{Berg}},
  \bibinfo{journal}{{Int. J. Frac. Mech.}} \textbf{\bibinfo{volume}{2}},
  \bibinfo{pages}{614} (\bibinfo{year}{1966}).

\bibitem[{\citenamefont{Tvergaard}(2008)}]{Tvergaard08}
\bibinfo{author}{\bibfnamefont{V.}~\bibnamefont{Tvergaard}},
  \bibinfo{journal}{Int. J. Mech. Sci.} \textbf{\bibinfo{volume}{50}},
  \bibinfo{pages}{1459} (\bibinfo{year}{2008}).

\bibitem[{\citenamefont{Tvergaard}(2009)}]{Tvergaard09}
\bibinfo{author}{\bibfnamefont{V.}~\bibnamefont{Tvergaard}},
  \bibinfo{journal}{Int. J. Frac.} \textbf{\bibinfo{volume}{158}},
  \bibinfo{pages}{41} (\bibinfo{year}{2009}).

\bibitem[{\citenamefont{Nielsen et~al.}(2012)\citenamefont{Nielsen, Dahl, and
  Tvergaard}}]{Nielsen12}
\bibinfo{author}{\bibfnamefont{K.~L.} \bibnamefont{Nielsen}},
  \bibinfo{author}{\bibfnamefont{J.}~\bibnamefont{Dahl}}, \bibnamefont{and}
  \bibinfo{author}{\bibfnamefont{V.}~\bibnamefont{Tvergaard}},
  \bibinfo{journal}{Int. J. Frac.} \textbf{\bibinfo{volume}{177}},
  \bibinfo{pages}{97} (\bibinfo{year}{2012}).

\bibitem[{\citenamefont{Benzerga et~al.}(2016)\citenamefont{Benzerga, Leblond,
  Needleman, and Tvergaard}}]{Benzerga16}
\bibinfo{author}{\bibfnamefont{A.~A.} \bibnamefont{Benzerga}},
  \bibinfo{author}{\bibfnamefont{J.-B.} \bibnamefont{Leblond}},
  \bibinfo{author}{\bibfnamefont{A.}~\bibnamefont{Needleman}},
  \bibnamefont{and}
  \bibinfo{author}{\bibfnamefont{V.}~\bibnamefont{Tvergaard}},
  \bibinfo{journal}{Int. J. Frac.} \textbf{\bibinfo{volume}{201}},
  \bibinfo{pages}{29} (\bibinfo{year}{2016}).

\bibitem[{\citenamefont{Morin et~al.}(2016)\citenamefont{Morin, Leblond, and
  Tvergaard}}]{Morin16}
\bibinfo{author}{\bibfnamefont{L.}~\bibnamefont{Morin}},
  \bibinfo{author}{\bibfnamefont{J.}~\bibnamefont{Leblond}}, \bibnamefont{and}
  \bibinfo{author}{\bibfnamefont{V.}~\bibnamefont{Tvergaard}},
  \bibinfo{journal}{J. Mech. Phys. Solids} \textbf{\bibinfo{volume}{94}},
  \bibinfo{pages}{148} (\bibinfo{year}{2016}).

\bibitem[{\citenamefont{Benzerga}(2002)}]{BenzergaJMPS}
\bibinfo{author}{\bibfnamefont{A.~A.} \bibnamefont{Benzerga}},
  \bibinfo{journal}{J. Mech. Phys. Solids} \textbf{\bibinfo{volume}{50}},
  \bibinfo{pages}{1331} (\bibinfo{year}{2002}).

\bibitem[{\citenamefont{Benzerga et~al.}(2004)\citenamefont{Benzerga, Besson,
  and Pineau}}]{Benzerga04AM-II}
\bibinfo{author}{\bibfnamefont{A.~A.} \bibnamefont{Benzerga}},
  \bibinfo{author}{\bibfnamefont{J.}~\bibnamefont{Besson}}, \bibnamefont{and}
  \bibinfo{author}{\bibfnamefont{A.}~\bibnamefont{Pineau}},
  \bibinfo{journal}{Acta Mater.} \textbf{\bibinfo{volume}{52}},
  \bibinfo{pages}{4639} (\bibinfo{year}{2004}).

\bibitem[{\citenamefont{{\student{Keralavarma}} and
  Benzerga}(2010)}]{KeralavarmaJMPS}
\bibinfo{author}{\bibfnamefont{S.~M.} \bibnamefont{{\student{Keralavarma}}}}
  \bibnamefont{and} \bibinfo{author}{\bibfnamefont{A.~A.}
  \bibnamefont{Benzerga}}, \bibinfo{journal}{J. Mech. Phys. Solids}
  \textbf{\bibinfo{volume}{58}}, \bibinfo{pages}{874} (\bibinfo{year}{2010}).

\bibitem[{\citenamefont{Benzerga and Leblond}(2014)}]{BenzergaJAM}
\bibinfo{author}{\bibfnamefont{A.~A.} \bibnamefont{Benzerga}} \bibnamefont{and}
  \bibinfo{author}{\bibfnamefont{J.-B.} \bibnamefont{Leblond}},
  \bibinfo{journal}{J. Appl. Mech.} \textbf{\bibinfo{volume}{81}},
  \bibinfo{pages}{031009} (\bibinfo{year}{2014}).

\bibitem[{\citenamefont{{\student{Torki}}
  et~al.}(2015)\citenamefont{{\student{Torki}}, Benzerga, and
  Leblond}}]{TorkiJAM}
\bibinfo{author}{\bibfnamefont{M.~E.} \bibnamefont{{\student{Torki}}}},
  \bibinfo{author}{\bibfnamefont{A.~A.} \bibnamefont{Benzerga}},
  \bibnamefont{and} \bibinfo{author}{\bibfnamefont{J.-B.}
  \bibnamefont{Leblond}}, \bibinfo{journal}{J. Appl. Mech.}
  \textbf{\bibinfo{volume}{82}}, \bibinfo{pages}{071005}
  (\bibinfo{year}{2015}).

\bibitem[{\citenamefont{Kweon et~al.}(2016)\citenamefont{Kweon, Sagsoy, and
  Benzerga}}]{KweonCMAME}
\bibinfo{author}{\bibfnamefont{S.}~\bibnamefont{Kweon}},
  \bibinfo{author}{\bibfnamefont{B.}~\bibnamefont{Sagsoy}}, \bibnamefont{and}
  \bibinfo{author}{\bibfnamefont{A.~A.} \bibnamefont{Benzerga}},
  \bibinfo{journal}{Comput. Methods Appl. Mech. Eng.}
  \textbf{\bibinfo{volume}{310}}, \bibinfo{pages}{495} (\bibinfo{year}{2016}).

\bibitem[{\citenamefont{Morin et~al.}(2015)\citenamefont{Morin, Leblond, and
  Benzerga}}]{MorinJMPS}
\bibinfo{author}{\bibfnamefont{L.}~\bibnamefont{Morin}},
  \bibinfo{author}{\bibfnamefont{J.-B.} \bibnamefont{Leblond}},
  \bibnamefont{and} \bibinfo{author}{\bibfnamefont{A.~A.}
  \bibnamefont{Benzerga}}, \bibinfo{journal}{J. Mech. Phys. Solids}
  \textbf{\bibinfo{volume}{75}}, \bibinfo{pages}{140} (\bibinfo{year}{2015}).

\bibitem[{\citenamefont{Eshelby}(1957)}]{Eshelby57}
\bibinfo{author}{\bibfnamefont{J.}~\bibnamefont{Eshelby}},
  \bibinfo{journal}{Proc. Roy. Soc} \textbf{\bibinfo{volume}{A241}},
  \bibinfo{pages}{357} (\bibinfo{year}{1957}).

\bibitem[{\citenamefont{Kailasam and Ponte~Castaneda}(1998)}]{Kailasam98}
\bibinfo{author}{\bibfnamefont{M.}~\bibnamefont{Kailasam}} \bibnamefont{and}
  \bibinfo{author}{\bibfnamefont{P.}~\bibnamefont{Ponte~Castaneda}},
  \bibinfo{journal}{J. Mech. Phys. Solids} \textbf{\bibinfo{volume}{46}},
  \bibinfo{pages}{427} (\bibinfo{year}{1998}).

\bibitem[{\citenamefont{Leblond and Mottet}(2008)}]{Leblond08}
\bibinfo{author}{\bibfnamefont{J.-B.} \bibnamefont{Leblond}} \bibnamefont{and}
  \bibinfo{author}{\bibfnamefont{G.}~\bibnamefont{Mottet}},
  \bibinfo{journal}{C. R. Mecanique} \textbf{\bibinfo{volume}{336}},
  \bibinfo{pages}{176} (\bibinfo{year}{2008}).

\bibitem[{\citenamefont{Shao et~al.}(2014)\citenamefont{Shao, Yang, Yao, and
  Liu}}]{Shao14}
\bibinfo{author}{\bibfnamefont{Y.}~\bibnamefont{Shao}},
  \bibinfo{author}{\bibfnamefont{G.-N.} \bibnamefont{Yang}},
  \bibinfo{author}{\bibfnamefont{K.-F.} \bibnamefont{Yao}}, \bibnamefont{and}
  \bibinfo{author}{\bibfnamefont{X.}~\bibnamefont{Liu}},
  \bibinfo{journal}{Appl. Phys. Lett.} \textbf{\bibinfo{volume}{105}},
  \bibinfo{pages}{181909} (\bibinfo{year}{2014}).

\end{thebibliography}

\end{document}